\documentclass[twocolumn]{webofc}
\DeclareUnicodeCharacter{03CA}{\"{\i}}
\usepackage[varg]{txfonts}   
\usepackage{hyperref}
\usepackage{url}
\hypersetup{colorlinks=true,citecolor=blue,urlcolor=blue,linkcolor=blue}
\usepackage{siunitx}
\DeclareSIUnit{\neutron}{neutron}
\usepackage[version=4]{mhchem}
\usepackage{graphicx}
\usepackage{subcaption}
\usepackage{caption}
\usepackage{lipsum}  
\usepackage{comment}

\usepackage{natbib}

\begin{document}
 
\title{Double-differential neutron-induced cross sections on Fe-nat at GANIL-NFS}
%
%

\author{\firstname{L.} \lastname{de Arruda}\inst{1,2,3} \and
        \firstname{D.} \lastname{Tarr\'io}\inst{2}\fnsep\thanks{Corresponding author; \email{diego.tarrio@physics.uu.se}} \and
        \firstname{X.} \lastname{Ledoux}\inst{1}\and
        \firstname{S.} \lastname{Pomp}\inst{2}\and
        \firstname{D.} \lastname{Ramos}\inst{1}\and
        \firstname{I.} \lastname{Ipatova}\inst{4}\and
        \firstname{P.} \lastname{Kanth}\inst{4}\and
        \firstname{I.} \lastname{Kodeli}\inst{4}\and
        \firstname{M.} \lastname{Lantz}\inst{2}\and
        \firstname{E.} \lastname{Pirovano}\inst{5}\and    
        \firstname{U.} \lastname{Tippawan}\inst{6}
}

\institute{
            Grand Accélérateur National d’Ions Lourds, CEA/DRF-CNRS/IN2P3, B.P. 55027, 14076, Caen, France 
\and
           Department of Physics and Astronomy, Uppsala University, Box 524, 75120 Uppsala, Sweden
\and
           Université de Caen-Normandie, Normandie Université, 14000 Caen, France
\and
           UKAEA, Culham Centre for Fusion Energy, Abingdon BOX14 3DB, United Kingdom
\and
           Physikalisch-Technische Bundesanstalt, D-38116 Braunschweig, Germany         
\and
           Department of Physics and Materials Science, Faculty of Science, Chiang Mai University, Chiang Mai, 50200, Thailand
}

\abstract{
An experimental campaign to measure double-differential cross sections (DDXS) for light-charged particles (LCPs) emitted from $^{\rm nat}$Fe was carried out in 2023 at the Neutrons For Science (NFS) facility of GANIL, France. The Medley experimental setup, specifically designed for light-ion measurements, was used for this campaign. The setup was installed in the white neutron beam line and irradiated with neutrons of energies up to \SI{44}{\mega\electronvolt}. The Time-of-Flight (TOF) technique was used to obtain $^{\rm nat}$Fe(n,pX) double-differential cross sections for all neutron energies within the white-neutron beam's energy range. This work highlights important steps in the data-analysis procedure and reports the first results from this experimental campaign. 
}
\maketitle
\section{Introduction}
\label{sec:intro}
Experimental nuclear data are critical for properly evaluating the damage to structural materials under strong radiation fields. Iron is the primary constituent of EUROFER steel \cite{eurofer}, for instance, and when subjected to high-energy neutron fields can suffer from radiation damage. Nuclear reactions induced by neutrons can produce light-charged particles (LCPs), i.e., isotopes of hydrogen and helium. These gases accumulate within the bulk material, leading to swelling and embrittlement. \cite{embrittlement}. 

Despite their relevance, there are very few sets of data well established above \SI{14}{\mega\electronvolt}. In this context, double-differential cross sections (DDXS) are particularly important, because they provide the energy and angular distributions of emitted particles, resulting in stricter constraints for nuclear models thus allowing more rigorous validation of evaluated libraries such as JEFF \cite{JEFF4} and ENDF \cite{ENDF}. This experiment addresses this gap, aiming to contribute with data for neutron energies up to \SI{44}{\mega\electronvolt}. In this work, we present the methods used to obtain such data together with preliminary measurements for \SI{14.5}{\mega\electronvolt} neutrons.

\section{The Medley setup at NFS}
\label{sec:setup}
The Medley setup was designed and built at Uppsala University, where it operated during several years measuring double differential cross sections for LCPs production \cite{Pomp_2010}. It is currently installed at the Neutrons for Science (NFS) facility in GANIL \cite{Ledoux_2021}, operating under a white neutron beam. A series of experiments for measuring DDXS for C-nat, Cr, and Fe-nat were carried out, and, more recently, a campaign for Cu-63, reported in \cite{Rasmusen2025}.

\begin{figure}[htb!]
\centering
\includegraphics[width=\columnwidth,clip]{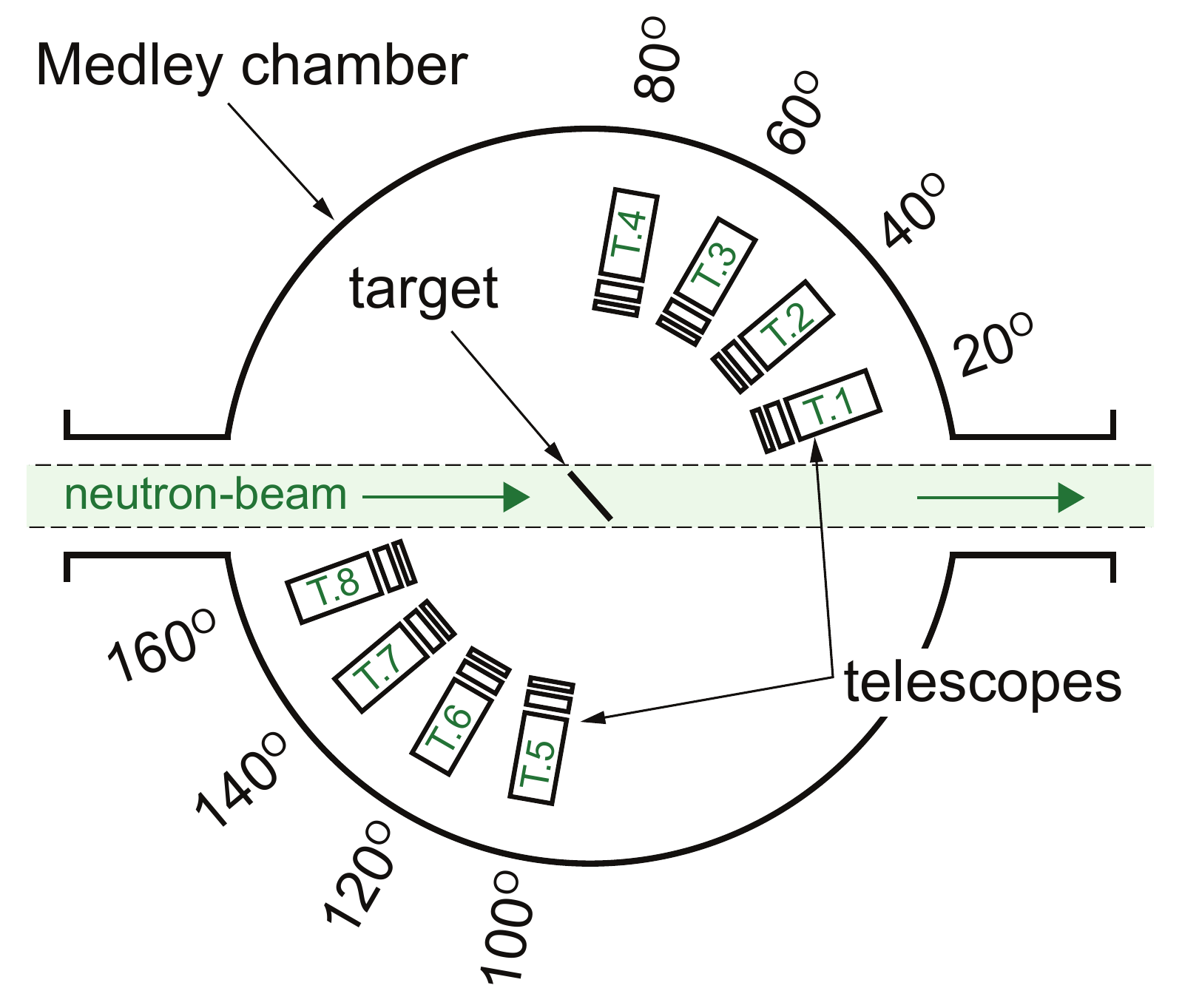}
\caption{Schematics of the Medley setup, showing the eight telescopes (T.1 to T.8), each comprising a sequence of three detectors, shown as rectangular areas. The distance between the telescopes and the target is approximately \SI{15}{\centi\metre}. The target is rotated \ang{45} with respect to the beam direction, to minimize energy losses of the produced LCP. The drawing is not to scale, for better visualization.}
\label{fig:medley_chamber}       
\end{figure}

The setup consists of a \SI{80}{\centi\metre} diameter vacuum chamber, where eight telescope detectors are installed around a central target holder, as shown in Fig.\,\ref{fig:medley_chamber}. Each telescope is composed of a stack of three detectors: two solid silicon depletion barrier (SSDB) followed by a CsI(Tl) scintillation detector. 

The telescopes are divided into two sets of four telescopes each (cf. Fig.\,\ref{fig:medley_chamber}): type `A' (tels. \numrange{1}{4}), equipped with silicon detectors of  \SI{50}{\micro\metre} and \SI{1000}{\micro\metre} thickness, respectively, and type `B' (tels. \numrange{5}{8}), with \SI{20}{\micro\metre} and \SI{500}{\micro\metre} thicknesses, respectively. The scintillation detectors are identical, having a thickness of about \SI{5}{\centi\metre}, as previously described in \cite{FluxdeArruda}. A Fe-nat target with an areal density of \SI{17.34(2)}{\milli\gram\per\centi\metre\squared} was used for the experiment reported in this work. 

\section{Particle identification and Time-of-Flight (TOF)}

The particle identification (PID) process is carried out with the aid of graphical cuts using $\rm \Delta E_{1}$,  $\rm \Delta E_{2}$ and  $\rm E_{res}$ energies, deposited in the first, second and third detector, respectively. The cuts select regions in the $\rm \Delta E_{1}:\Delta E_{2}$ and $\rm \Delta E_{2}: E_{res}$ plots. The telescopes present good energy resolution making it possible to distinguish the hydrogen and helium isotopes (p, d, t, and  He-3, and $\alpha$, respectively). 

\begin{figure}[htb!]
\centering
\begin{subfigure}[b]{\columnwidth}
    \centering
    \includegraphics[width=\linewidth,clip]{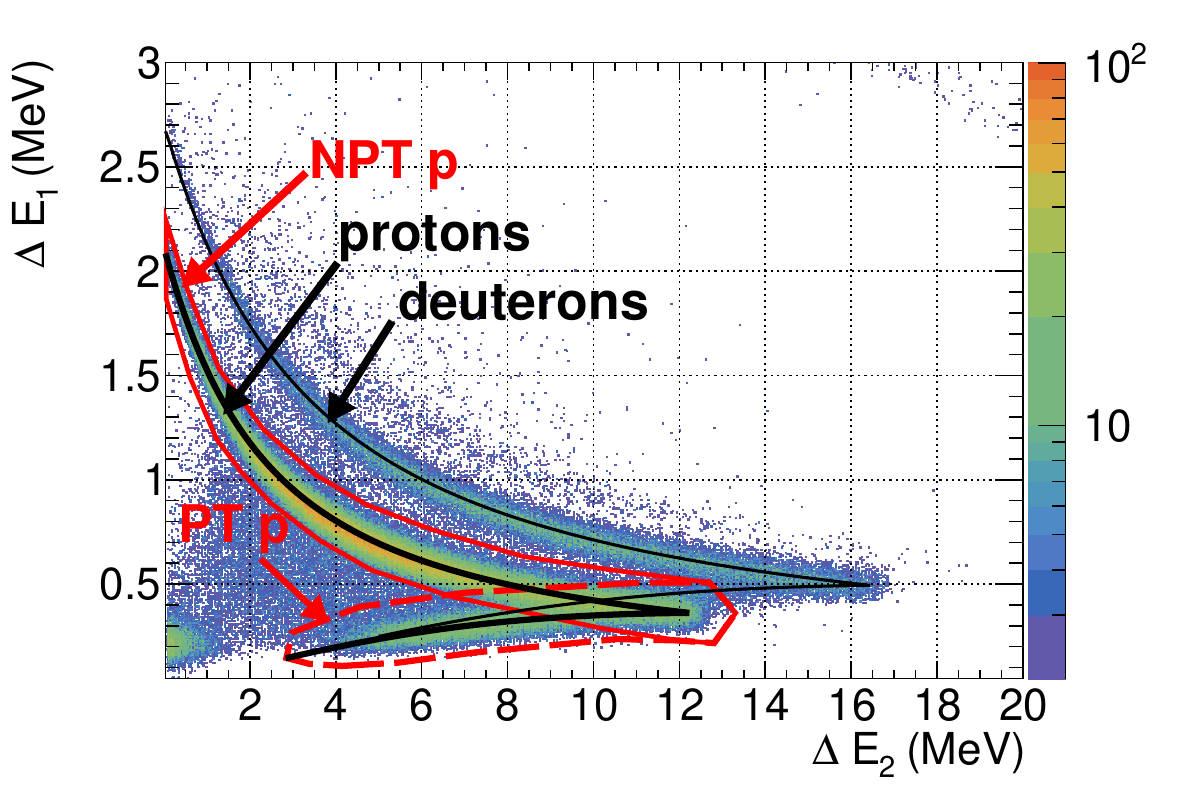}
    \caption{}
    \label{fig:PIDa}
\end{subfigure}

\vspace{1em}  

\begin{subfigure}[b]{\columnwidth}
    \centering
    \includegraphics[width=\linewidth,clip]{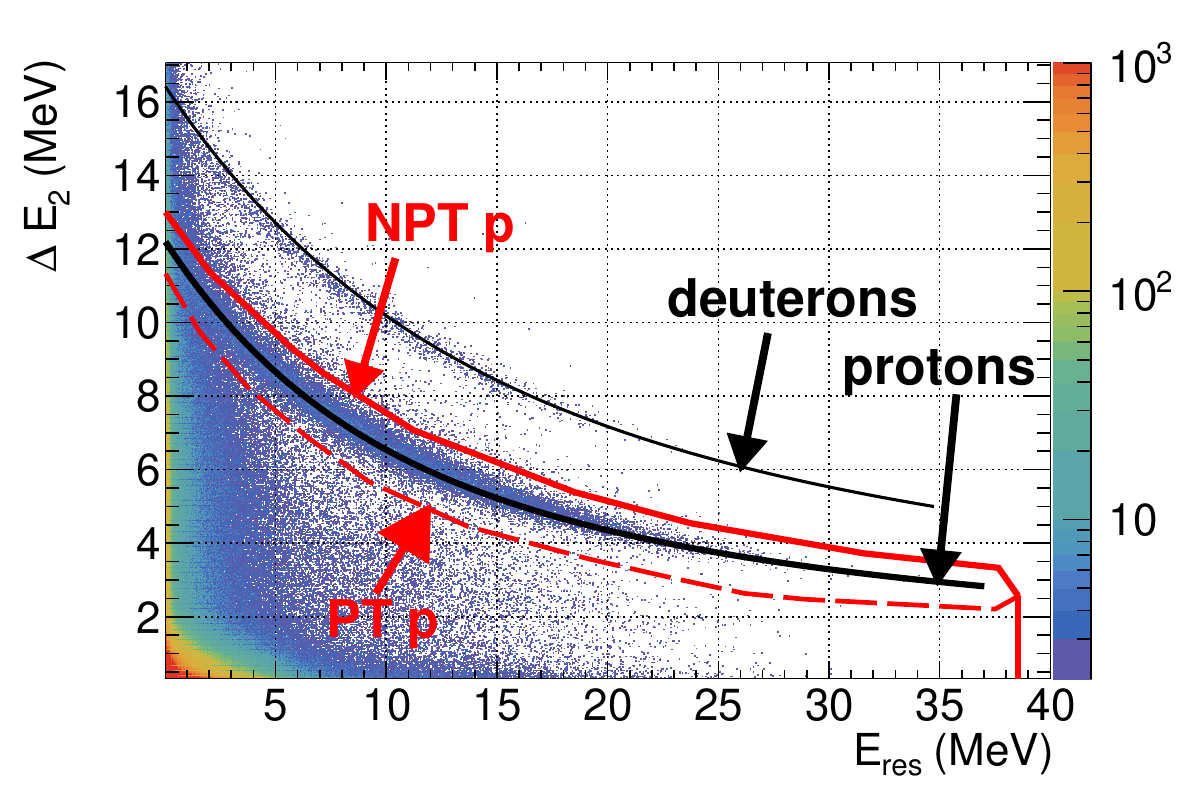}
    \caption{}
    \label{fig:PIDb}
\end{subfigure}
\caption{Particle identification process: (a) $\rm \Delta E_{1}:\Delta E_{2}$ plot, with the protons and deuterons identified through the plotted calculations (solid black lines).  The graphical cuts for non-punched-through (`NPT p') and punched-through (`PT p') protons are also shown. (b) $\rm \Delta E_{2}: E_{res}$ plot with the analogous graphical cuts.}
\label{fig:PID}
\end{figure}

This study considers only the proton case, for which the corresponding cuts are presented in Fig.\,\ref{fig:PID}, together with the energy-loss calculations, shown as solid curves. The LCPs are divided into non-punching-through (NPT) or punching-through (PT), depending on whether the LCP punched-through the second silicon detector. The corresponding cuts are labelled accordingly in both plots of Fig.\,\ref{fig:PID}.

The neutron energies are determined via the Time-of-Flight (TOF) method. The time for an event is measured in the second silicon detector of each telescope, and is defined as the difference between the detection of the LCP and the next radio-frequency (RF) signal of the NFS linear accelerator (LINAC), which is in phase with the neutron bunches. The second silicon detectors are sensitive to the gamma flash produced together with the neutrons, despite their lower gamma-detection efficiency. As such gammas are synchronized with the RF signals, they are used as a time signature for neutron production, making it possible to retrieve the neutron TOF on an event-by-event basis. The time resolution of the setup is \SI{6}{\nano\second}. 

A distortion in the TOF measurement was identified as a consequence of the different shapes of the LCP signals, depending on its energy \cite{Tarrio2026_timing}. This was corrected by adjusting an energy dependent time-shift, as explained in Ref. \cite{FluxdeArruda}. 

\section{Thick-target correction}
\label{sec:TTC}

A thick target correction (TTC) must be applied to account for deformations in the measured spectrum caused by energy losses within the target. To compute such a correction, energy losses were simulated using Geant4 \cite{geant4} to obtain a response function, which was then used to reconstruct the expected initial energy $\rm E_0$ of a detected LCP from its measured energy $\rm E'$. 

\begin{figure}[htb!]
\centering
\includegraphics[width=\columnwidth,clip]{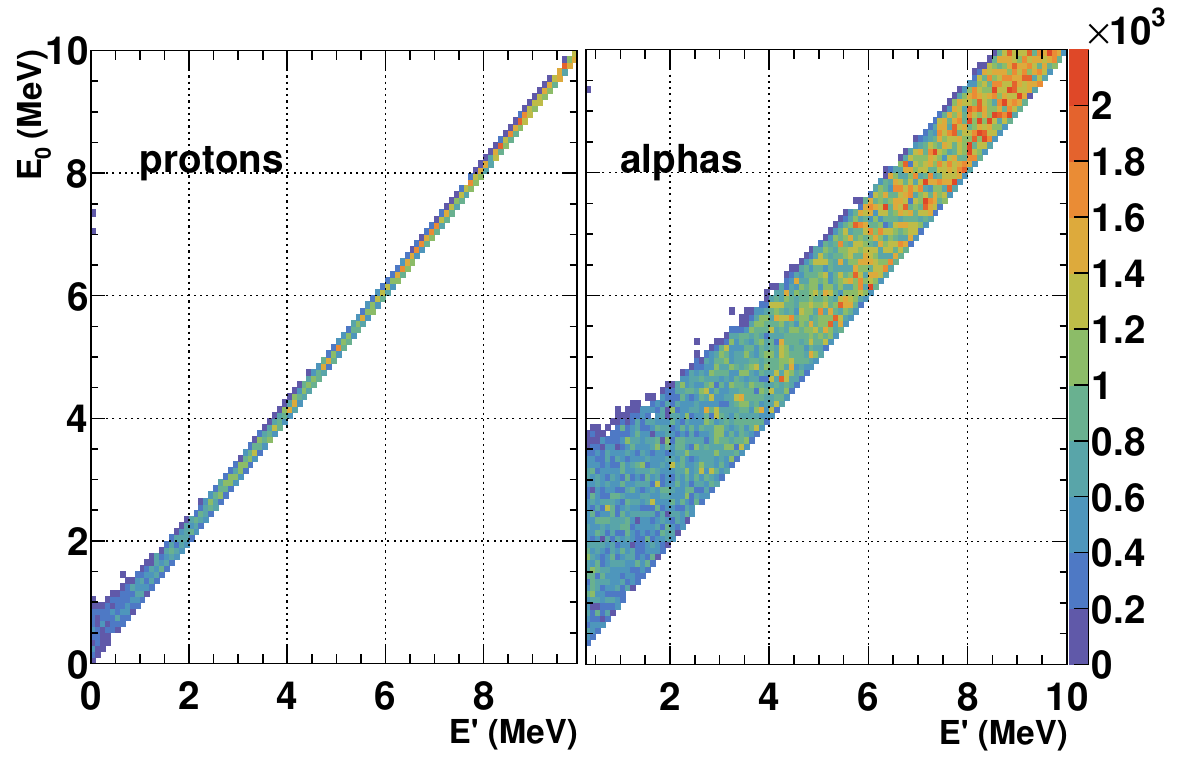}
\caption{Simulated response matrices for protons and alphas emitted from the target and being detected in the telescope at \ang{20}. The matrices cover up to \SI{44}{\mega\electronvolt}, but the plots show a zoomed range.}
\label{fig:Rmatrices}       
\end{figure}

The simulation consists of the target geometry (within which the particles are uniformly generated) together with the geometries for the first silicon detector of each telescope. The particles are then emitted from its creation point within the target, eventually hitting any of the detectors. A flat initial energy spectrum distribution in the \mbox{\qtyrange[range-units = single]{0}{44}{\mega\electronvolt}} range is used. From the results, one can retrieve the response functions for each particle and each telescope, as shown in Fig.\,\ref{fig:Rmatrices}, for the telescope placed at \ang{20}. It is also possible to account for the number of particles lost within the target, that has to be considered for a proper deconvolution. 

To reduce the dependence of the unfolded results on the flat prior spectrum assumed in the response matrix calculation, an iterative Bayesian approach using the well-established D'Agostini method \cite{DAGOSTINI1995} was applied. In this method, the measured data sequentially corrects the estimated true distribution. The RooUnfold implementation of this method was used \cite{roounfold}. 

A validation of the method was carried out, and also used to find the optimal number of iterations needed for a correct reconstruction. For this task, the proton emission spectrum at \ang{20} in the laboratory reference frame was calculated using the TALYS code \cite{Koning2023} for several neutron energies, and then inserted in the Geant4 simulation, calculating the degraded proton spectrum detected by telescope T.1. Subsequently, the response function of Fig.\,\ref{fig:Rmatrices} was used in the D'Agostini method to obtain the reconstructed spectrum, for several iterations of the method. At each iteration, the squared differences between the original and reconstructed spectra were recorded. The optimal number of iterations was three in all cases, since it is the number of iterations that minimizes the difference between the original and reconstructed distributions.

\begin{figure}[htb!]
\centering
\includegraphics[width=\columnwidth,clip]{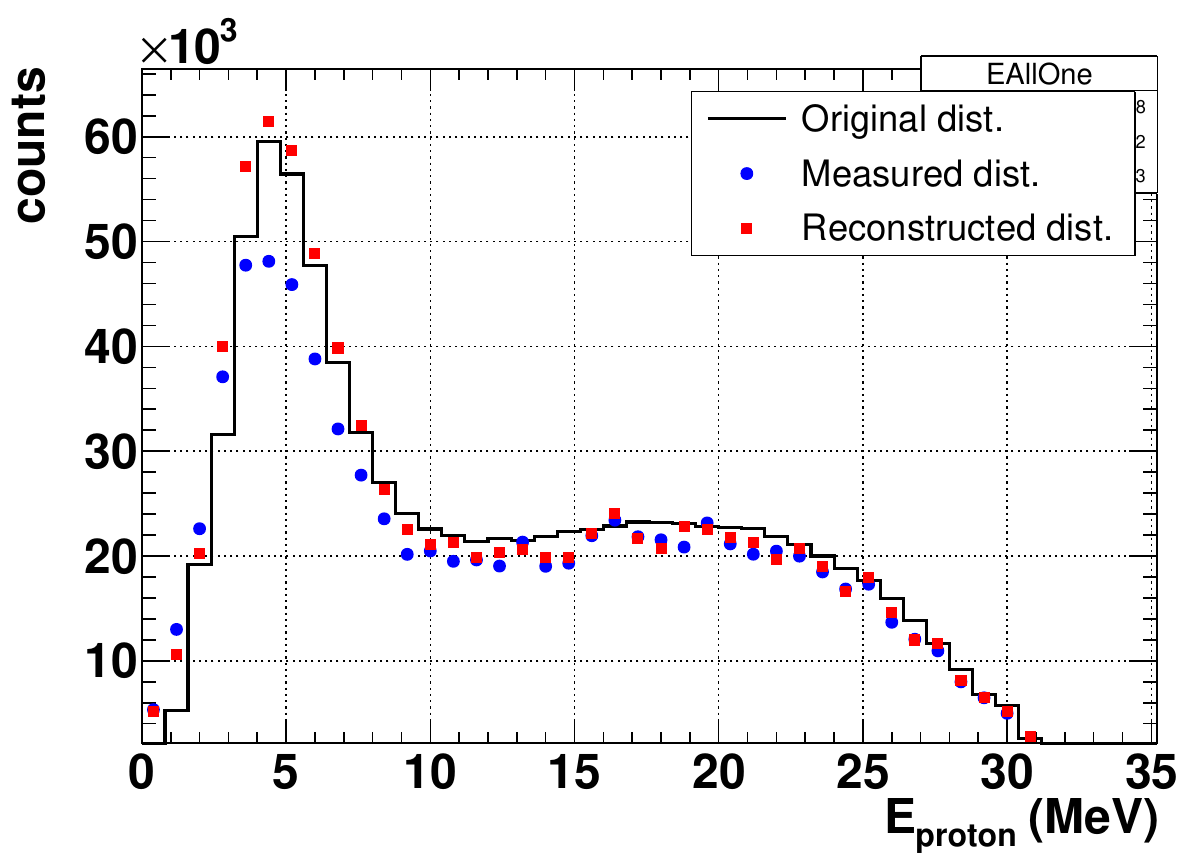}
\caption{Validation of the deconvolution method showing the original \ang{20} emitted protons distribution (solid line) calculated by TALYS code, the simulated measured distribution at telescope T.1, and the reconstruction of the original expected distribution using the response matrix calculated with flat prior (from the left panel of Fig.\,\ref{fig:Rmatrices}). The reconstruction was carried out using D'Agostini method \cite{DAGOSTINI1995}, with \num{3} iterations.}
\label{fig:reco}       
\end{figure}

\section{Double-Differential cross sections}
\label{sec:DDX}

\begin{figure}[htb!]
\centering
\includegraphics[width=\columnwidth,clip]{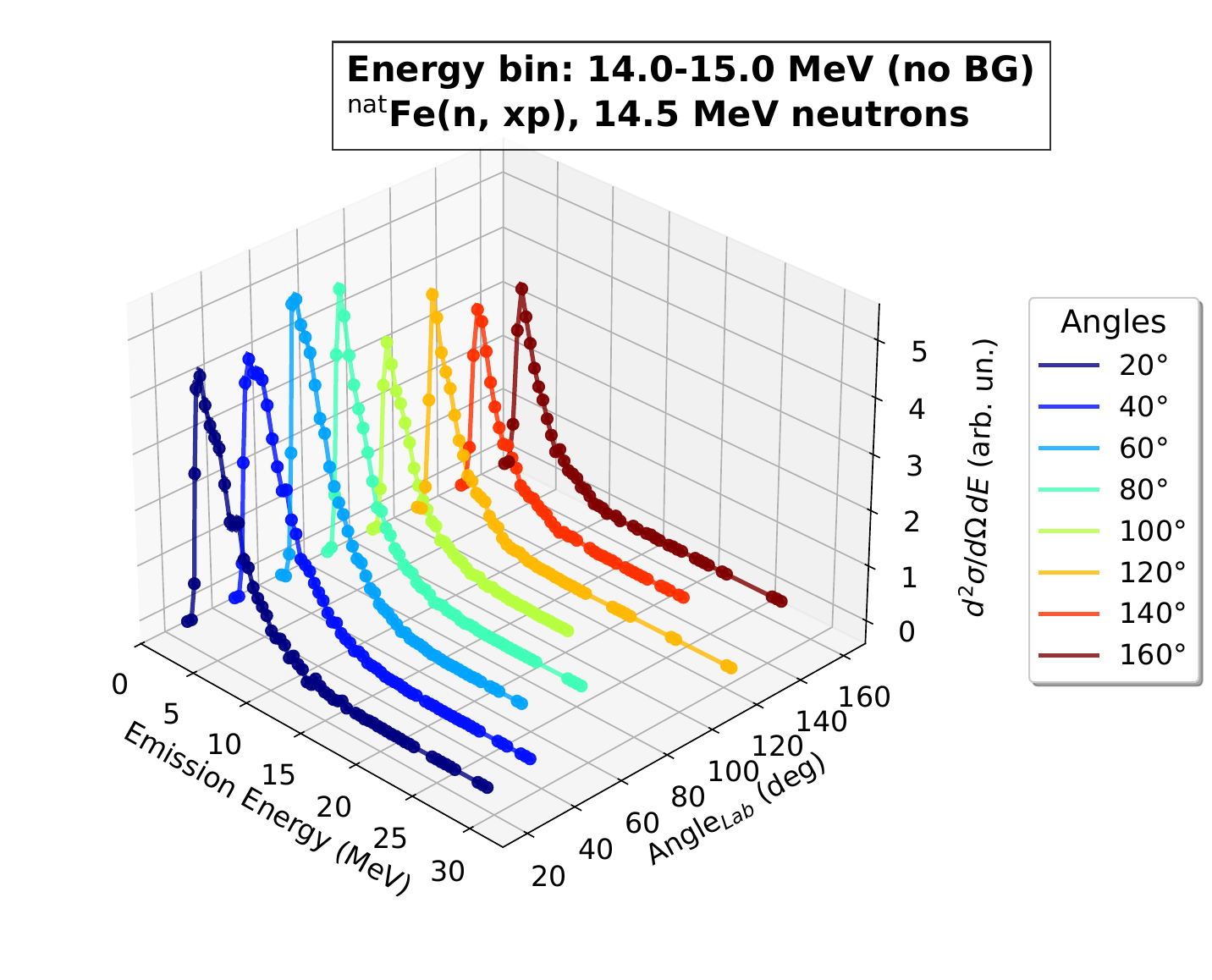}
\caption{Example of preliminary DDXS measured for Fe-nat in the 2023 Medley campaign, for \SI{14.5}{\mega\electronvolt} neutrons.}
\label{fig:DDX}       
\end{figure}

The Double-differential cross sections (DDXS) can be retrieved, for each neutron energy interval $\rm \Delta E_{NN}$ as:

\begin{equation}
    \rm \frac{d^2\sigma}{d\Omega dE } (E,\rm \theta_{Lab}) = \frac{N_{\text{counts}} \cdot L^{2}}{\phi \cdot \Delta E_{NN} \cdot \Omega_{\text{det}} \cdot \Delta E_{p} \cdot Q \cdot N_{\text{atoms}}},
\end{equation}

where $\rm N_{counts}$ is the number of LCPs (with energy $\rm E$, detected at $\rm \theta_{Lab}$), $\rm L$ is the distance between the NFS' neutron source to the target (centred in the Medley chamber), $\phi$ is the neutron flux given in \unit{\neutron\per\steradian\per\mega\electronvolt\per\micro\coulomb}, $\Omega_{\rm det}$ is the solid angle of the detector under analysis, $ \rm Q$ is the total deuteron charge used to produce the neutrons for the beam, and $\rm N_{atoms}$ is the total number of Fe nuclei in the target. After subtracting the background measurements, we could retrieve preliminary DDXS for different energies, as the example shown in Fig.\,\ref{fig:DDX}. Some fine tuning is still ongoing.

\begin{figure}[htb!]
\centering
\includegraphics[width=\columnwidth,clip]{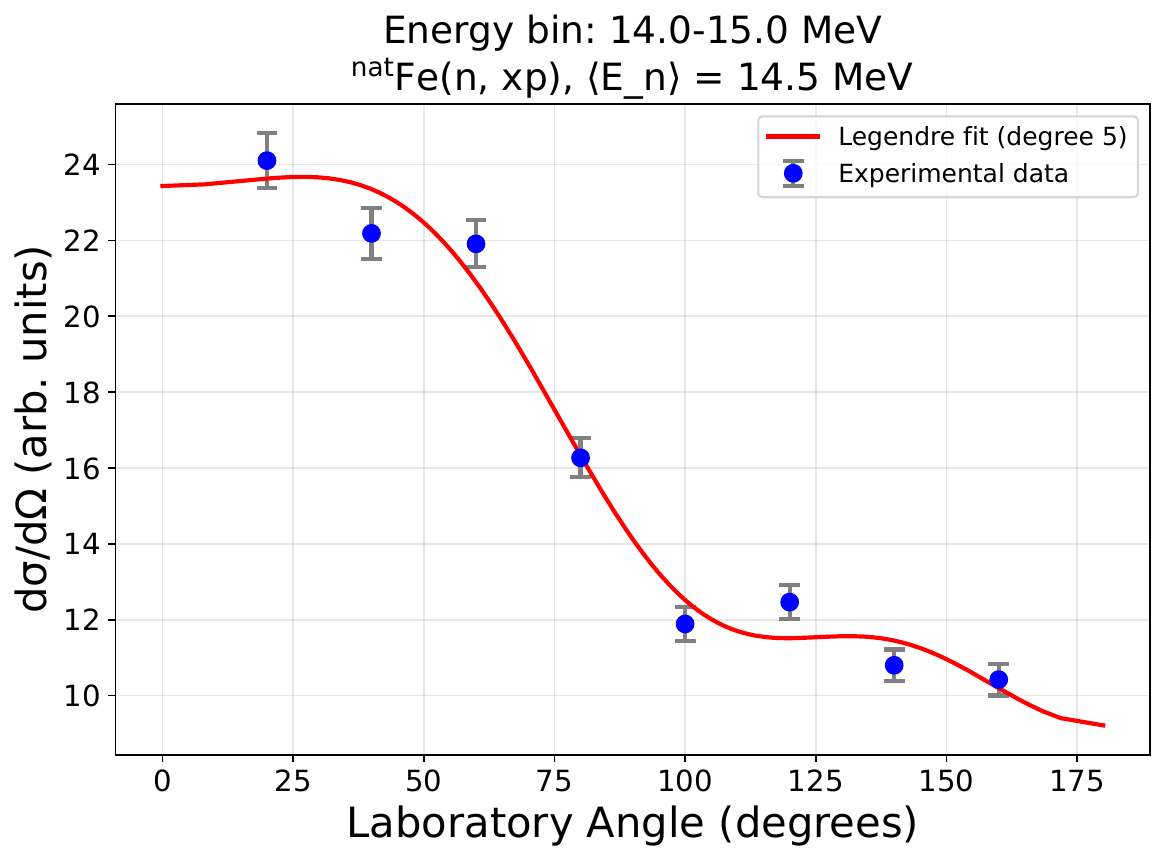}
\caption{Preliminary angular differential cross-sections as a function of the angle in the laboratory reference frame, for \SI{14.5}{\mega\electronvolt} neutrons.}
\label{fig:DX}       
\end{figure}

By integrating the DDXS along all the proton emission energies, one can obtain the angular differential cross sections (DXS), as a function of the production angle in the laboratory reference frame. By choosing the Legendre's polynomials as a base for fitting the distributions, as exemplified in Fig.\,\ref{fig:DX}, one can also obtain the total production cross sections via integration.  

\section{Summary and outlook}
\label{sec:conclusions}
This work presents preliminary Fe-nat(n,pX) cross section results, for the Fe-nat experimental campaign carried out at NFS, using the Medley setup operating under a white beam. Several technical challenges were identified and solved. The setup proved to properly work in the facility, with the white-neutron beam.

The TTC method has proven to work well, and good fits could also be obtained for the angular distributions. Analysis for the total production cross sections is ongoing. Targets with different thicknesses were available for this experimental campaign, and will allow to study better the different energy ranges for the other LCPs.

\section{Acknowledgements}

The authors would like to acknowledge Johan Goupil, Patrice Bourgault, Jean-Claude Foy, and the GANIL local team, for their valuable contributions to this experiment. This work has received support from the European Union’s research and training program under grant agreements with numbers 847594 (ARIEL), and 101057511 (EURO-LABS). Part of this work has been carried out within the framework of the EURO-fusion Consortium, funded by the European Union via the Euratom Research and Training Programme (Grant Agreement No 101052200 — EUROfusion). This work was also supported by the Swedish Radiation Authority (SSM) under grant agreement SSM2024-1205-17, by the MÅSTE project (P2023-01281) funded by the Swedish Energy Agency, and by the Swedish Center for Nuclear Technology (SKC).



\bibliography{references}
\end{document}